\documentclass[12pt,aps,prl,draft]{revtex4}

\usepackage{epsf}
\usepackage{bm}

\begin{document}
\title{Finite-thickness effects in plasmonic films with periodic cylindrical anisotropy}

\author{Igor V.~Bondarev}\email[E-mail: ]{ibondarev@nccu.edu}
\affiliation{Math \& Physics Department, North Carolina Central University, Durham, NC 27707, USA}

\begin{abstract}
Finite-thickness effects are analyzed theoretically for the plasma frequency and associated dielectric response function of plasmonic films formed by periodically aligned, infinitely thin, identical metallic cylinders. The plasma frequency of the system is shown to have the unidirectional square-root-of-momentum and quasi-linear momentum spatial dispersion for the thick and ultrathin films, respectively. This spatial dispersion and the unidirectional dielectric response nonlocality associated with it can be adjusted not only by the film material composition but also by varying the film thickness, the cylinder length, the cylinder-radius-to-film-thickness ratio, and by choosing the substrates and superstrates of the film appropriately. Application of the theory developed to the finite-thickness periodically aligned carbon nanotube films is discussed.
\end{abstract}

\maketitle

\section{Introduction}

Plasmonic films are the major components used to create metamaterial structures with reduced dimensionality (quasi-2D) --- optical metasurfaces~\cite{KBSh13}. Metamaterials are usually highly anisotropic metal-dielectric composites, which may be fabricated to have metallic response in one direction and dielectric response in the other directions~\cite{Pendry06,Shalaev07,CaiShalaev,Brongersma12}. This enables many exotic applications such as nanoscale imaging, efficient light concentration~\cite{Brongersma12,Zubin06,Zhang07,Engheta06,Podolskiy07,Atwater10,Sasha11,Halas11} and, most importantly, the capability of the photonic density of states engineering to be able to control near-surface radiative and nonradiative processes~\cite{Menon12,Zubin12,Noginov10}. With an extra feature of reduced dimensionality, optical metasurfaces offer new exceptional abilities of controlling incident electromagnetic flow which open up the door to the applications such as single-photon sources with \emph{directionally} increased photon extraction useful for quantum information technologies and microscopy, for imaging and sensing as well as for probing the fundamentals of the light-matter interactions at the nanoscale~\cite{Koppens11,Koppens14,Basov15,Rodrigo,Mak,Stauber17}. A key to realizing these applications is the ability to fabricate ultrathin metallic films of precisely controlled in-plane anisotropy, periodicity and thickness, which is quite possible with the modern progress in nanofabrication techniques~\cite{Hecht10,Kono16NatNano,Shalaevgroup17,Shalaevgroup17ACS,Shalaevgroup18}. However, the strong vertical electron confinement entails new quantum effects~\cite{GarciaAbajo14,BondarevShalaev17,BondarevShalaev18MRS} which are still to be explored for their importance in controlling the optical properties of ultrathin plasmonic films~\cite{BondarevShalaev18ACES}, especially in the presence of the in-plane anisotropy.

Very recently, Bondarev and Shalaev have proposed a rigorous theoretical model to account for the vertical (out-of-plane) electron confinement effect on the in-plane electron plasma oscillations in ultrathin metallic films of finite thickness~\cite{BondarevShalaev17}. The model uses the \emph{isotropic} 2D-momentum ($k$) Fourier transform of the pair Coulomb interaction potential in the Keldysh-Rytova form~\cite{Keldysh79} to obtain the classical Lagrange equations of motion for the electron gas density in the film. The thin film plasma frequency thus obtained, while being independent of $k$ for relatively thick films, acquires the $\sqrt{k}$ spatial dispersion typical of 2D materials~\cite{Basov} and gradually shifts to the red as the film thickness decreases. This agrees with and thereby explains the recent plasma frequency measurements done on stoichiometrically perfect ultrathin TiN films of controlled variable thickness~\cite{Shalaevgroup17,Shalaevgroup18}. This theory was then used to demonstrate the remarkable confinement-induced features of the ultrathin films of finite lateral size~\cite{BondarevShalaev18MRS}. They are the resonance magnetic response and the low-frequency negative refraction resulted from the plasma frequency spatial dispersion and associated \emph{nonlocality} of the dielectric response of the film.

In this article, the same approach is used to derive and analyze the plasma frequency and associated dielectric response function of the finite-thickness plasmonic film formed by an array of periodically aligned, infinitely thin, identical metallic cylinders. The key features that make this system interesting are the periodic cylindrical alignment and the spatially periodic anisotropy associated with it. Thin films of aligned carbon nanotubes (CNs) present an important example of such systems with periodic cylindrical anisotropy. CN array and related nanotube superlattice systems have been in the process of intensive experimental development in the hope of creating a new generation of multifunctional ultrathin metasurfaces and nonlinear optical devices with characteristics adjustable on demand by means of the CN diameter, chirality and periodicity variation~\cite{Kono16NatNano,Abram17NL,Abram17PRL}.

\begin{figure}[t]
\epsfxsize=14.cm\centering{\epsfbox{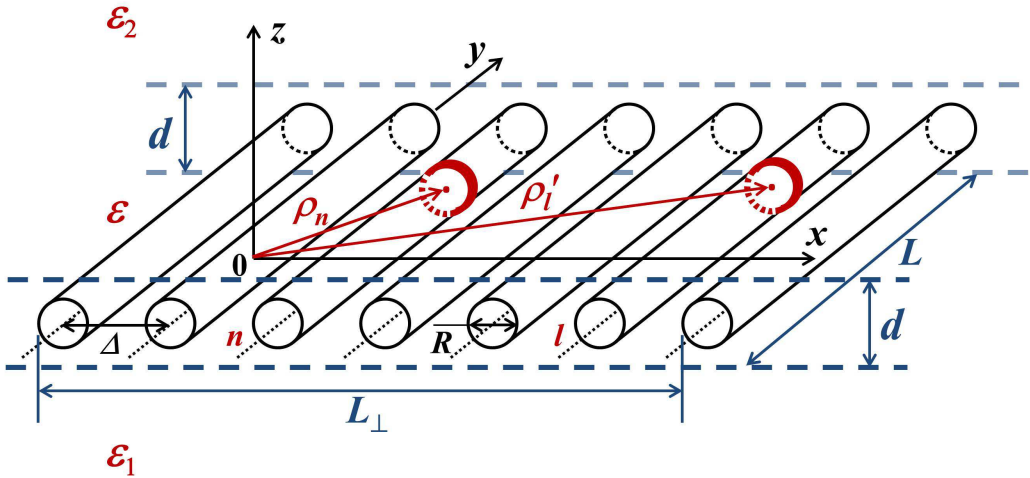}}
\caption{Schematic to show the geometry notations for the model of the finite-thickness plasmonic film with periodic cylindrical anisotropy. See text for details.}\label{fig1}
\end{figure}

\section{The model}

The model system under study here is presented in Fig.~\ref{fig1}. The periodic array (the film) of parallel, infinitely thin, identical metallic cylinders of radius $R$ and length $L$ has the translational unit $\Delta$, width $L_\perp$ and thickness $d$. The array is immersed in a dielectric to make the film have the \emph{effective} dielectric constant $\varepsilon$. The film is sandwiched between the substrate and the superstrate with the dielectric constants $\varepsilon_1$ and $\varepsilon_2$, respectively. The array parameters are assumed to obey the set of constraints $2R\!\le\!\Delta\!\le\!d\!\ll\!L\!\sim\!L_\perp$ with $R$ and $\Delta$ being much less than the wavelength of light radiation.

When $\varepsilon\!\gg\!\varepsilon_1+\varepsilon_2$, which is assumed to be the case here as well, and $d$ is much less than the typical in-plane ($x,y$) distance between the charges confined in the film, the increased 'outside' contribution to the pair Coulomb interaction energy makes the Coulomb interaction between the charges confined stronger than that in a homogeneous medium with the dielectric constant $\varepsilon$, while the pair Coulomb interaction potential loses its out-of-plane ($z$) coordinate dependence to turn into a pure in-plane 2D potential~\cite{Keldysh79}. In the case here, one has the electron charge density constrained to be distributed all over the periodic cylindrical surfaces homogeneously, whereby the pair electron Coulomb interaction can be approximated by that of two uniformly charged rings of radius $R$~\cite{Louie09}, embedded in a dielectric layer of thickness $d$. For two such rings of the unit cells at points $\bm{\rho}_n$ and $\bm{\rho}_l^{\,\prime}$ of cylinders $n$ and $l$ as shown in Fig.~\ref{fig1}, the Coulomb interaction energy can be expanded in the reciprocal (2D-momentum) space to take the following form~\cite{Louie09,BondarevShalaev17}
\begin{equation}
V(\bm{\rho}_n-\bm{\rho}^{\,\prime}_l)\approx\frac{8\pi e^2}{\varepsilon LL_\perp}\sum_{\mathbf{k}}~^{\!\!\!\textstyle^\prime}
\frac{qRI_0(qR)K_0(qR)}{k[kd+(\varepsilon_1+\varepsilon_2)/\varepsilon]}\exp\left[i\mathbf{k}\!\cdot\!(\bm{\rho}_n-\bm{\rho}^{\,\prime}_l)\right]\,.
\label{KeldyshFourier}
\end{equation}
Here, $I_0$ and $K_0$ are the zeroth-order modified cylindrical Bessel functions, $\mathbf{k}\!=\!\mathbf{q}+\mathbf{k}_\perp$ is the in-plane electron quasimomentum with $\mathbf{q}$ and $\mathbf{k}_\perp$ representing its components in the $y$- and $x$-direction --- parallel and perpendicular to the cylinder alignment direction, respectively, $k\!=\!|\mathbf{k}|\!=\!\sqrt{q^2+k_\perp^2}$ and $k_\perp\!=\!2\pi n_{x}/L_\perp$, where $n_{x}\!=\!0,\pm1,\pm2,...,\pm N/2$ with $N$ being the total number of cylinders so that $|k_\perp|\!\le\!\pi/\Delta$. The summation sign is primed to indicate that the term with $\mathbf{k}\!=\!0$ associated with the all-together electron displacement must be dropped and $\bm{\rho}_n\!\ne\!\bm{\rho}_l^{\,\prime}$ to exclude the self-interaction.

\section{The plasma frequency and optical response}

With Eq.~(\ref{KeldyshFourier}) the Coulomb potential energy of the charged ring of the unit cell at point $\bm{\rho}_n$ takes the form
\begin{equation}
V(\bm{\rho}_n)=\frac{8\pi e^2}{\varepsilon LL_\perp}\sum_{l,\mathbf{k}}~^{\!\!\!\textstyle^\prime}
\frac{qRI_0(qR)K_0(qR)}{k[kd+(\varepsilon_1+\varepsilon_2)/\varepsilon]}\exp\left[i\mathbf{k}\!\cdot\!(\bm{\rho}_n-\bm{\rho}^{\,\prime}_l)\right],
\label{poten}
\end{equation}
where the $l$-summation runs over the different cylinders and over their individual unit cells to also include the unit cells of cylinder $n$ but the cell at point $\bm{\rho}_n$ itself. Along with the electron kinetic energy $K\!=\!\sum_lm^\ast\dot{\bm\rho}_l^{\,2}/2$, where $m^\ast$ is the electron effective mass, one then arrives at the equations of motion
\begin{equation}
\ddot{\bm\rho}_n=\frac{8\pi e^2}{\varepsilon m^\ast LL_\perp}\sum_{l,\mathbf{k}}~^{\!\!\!\textstyle^\prime}
\!(-i\textbf{k})\,\frac{qRI_0(qR)K_0(qR)}{k[kd+(\varepsilon_1+\varepsilon_2)/\varepsilon]}
\exp\left[i\mathbf{k}\!\cdot\!(\bm{\rho}_n-\bm{\rho}^{\,\prime}_l)\right].
\label{acceleration}
\end{equation}

Introducing the local electron density
\[
n(\bm\rho)=\sum_l\delta(\bm\rho-\bm\rho_l)=\sum_\mathbf{k}n_\mathbf{k}\exp\left(i\mathbf{k}\!\cdot\!\bm\rho\right)\nonumber\\
\]
with the Fourier components
\begin{equation}
n_\mathbf{k}=\frac{1}{LL_\perp}\sum_l\exp\left(-i\mathbf{k}\!\cdot\!\bm\rho_l\right),\;\;\;n_{\mathbf{k}=0}=N_{2D}\hskip0.5cm
\label{N2D}
\end{equation}
($N_{2D}$ being the equilibrium \emph{surface} electron density) and using Eq.~(\ref{acceleration}), one obtains
\[
\ddot{n}_\mathbf{k}=-\frac{8\pi e^2}{\varepsilon m^\ast}\sum_{\mathbf{k}^\prime}~^{\!\!\!\textstyle^\prime}
\frac{\left(\textbf{k}\!\cdot\!\mathbf{k}{'}\right)n_{\mathbf{k}{'}}n_{\mathbf{k}-\mathbf{k}{'}}q{'\!}RI_0(q{'\!}R)K_0(q{'\!}R)}
{k{'}[k{'}d+(\varepsilon_1+\varepsilon_2)/\varepsilon]}
-\frac{1}{LL_\perp}\sum_l\left(\textbf{k}\!\cdot\!\dot{\bm\rho}_l\right)^2\exp\left(-i\mathbf{k}\!\cdot\!\bm\rho_l\right).
\]
This can now be simplified in the random phase approximation (RPA) by dropping the alternating-sign terms ($\mathbf{k}{'}\!\ne\!\mathbf{k}$) in the sum over $\mathbf{k}{'}$ (see Ref.\cite{Pines} for details about the RPA validity range), to obtain after using $N_{2D}$ of Eq.~(\ref{N2D}) the expression as follows
\begin{eqnarray}
\ddot{n}_\mathbf{k}+\omega_p^2(k)\,n_\mathbf{k}=-\frac{1}{LL_\perp}\sum_l\left(\textbf{k}\!\cdot\!\dot{\bm\rho}_l\right)^2\exp\left(-i\mathbf{k}\!\cdot\!\bm\rho_l\right)
\label{maineqnrpa}
\end{eqnarray}
with
\begin{equation}
\omega_p(k)=\omega_p(q,k_\perp)=\sqrt{\frac{8\pi e^2N_{2D}\,qRI_0(qR)K_0(qR)}{\varepsilon m^\ast d\,[1+(\varepsilon_1+\varepsilon_2)/\varepsilon kd]}}\,.
\label{omegapofk}
\end{equation}
This turns into the harmonic oscillator equation given $\mathbf{k}^2\!\ll\!\mathbf{k}_c^2\!=\!\omega_p^2/v_0^2$ with~$v_0$ defined by $m^\ast v_0^2/2\!=\!E_F$ (or $k_BT$) for the degenerate (non-degenerate) electron gas system~\cite{Pines}. When $\mathbf{k}$ ($=\!\mathbf{q}\!+\!\mathbf{k}_\perp$) is much less than the cut-off vector $\mathbf{k}_c$, the right hand side of Eq.~(\ref{maineqnrpa}) becomes much less than $\mathbf{k}_c^2v_0^2\,n_{\mathbf{k}}\!=\!\omega_p^2\,n_{\mathbf{k}}$ to result in the collective electron density oscillations --- plasma oscillations --- with the plasma frequency featuring \emph{anisotropic} spatial dispersion as given by Eq.~(\ref{omegapofk}).

\begin{figure}[t]
\epsfxsize=9.3cm\centering{\epsfbox{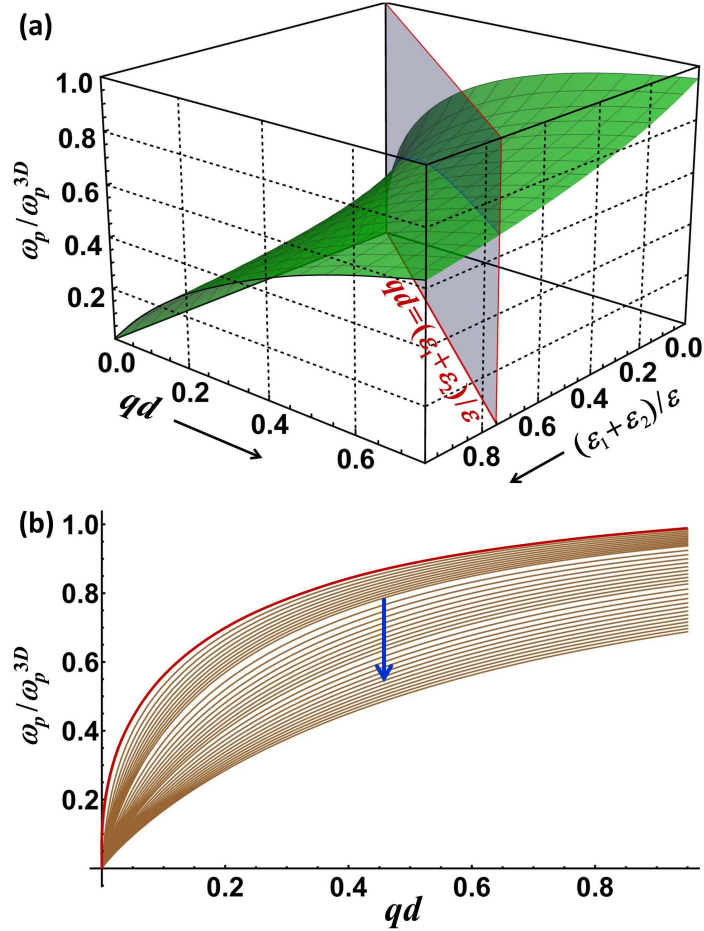}}
\caption{(a)~The ratio $\omega_p/\omega_p^{3D}$ given by Eq.~(\ref{omegapofq}) with $R/d\!=\!0.5$ as a function of the dimensionless variables $qd$ and $(\varepsilon_1+\varepsilon_2)/\varepsilon$. ~(b)~The contour plot obtained by cutting the graph in (a) with parallel vertical planes of constant $(\varepsilon_1+\varepsilon_2)/\varepsilon$. The vertical blue arrow shows the direction of the $(\varepsilon_1+\varepsilon_2)/\varepsilon$ increase.}
\label{fig2}
\end{figure}

One can see that $\omega_p(0,k_\perp)=\omega_p(k_\perp)=0$ due to the properties of the modified cylindrical Bessel functions, while $\omega_p(q,0)=\omega_p(q)\ne0$ and is strongly thickness dependent. No plasma oscillations occur and the film behaves as a dielectric in the direction perpendicular to the cylinder alignment ($x$-direction in Fig.~\ref{fig1}). There are plasma oscillations in the cylinder alignment direction ($y$-direction in Fig.~\ref{fig1}), in which case the plasma frequency takes the form
\begin{equation}
\omega_p(q)=\omega_p^{3D}\sqrt{\frac{2qRI_0(qR)K_0(qR)}{1+(\varepsilon_1+\varepsilon_2)/\varepsilon qd}}\,,
\label{omegapofq}
\end{equation}
where $\omega_p^{3D}\!=\!\sqrt{4\pi e^2N_{3D}/\varepsilon m^\ast}$ is the \emph{effective} bulk plasma frequency of the film material ($N_{3D}\!=\!N_{2D}/d$ being the \emph{volumetric} electron density), whereby the film behaves as a spatially dispersive metal with the dispersion character controlled by the thickness $d$ and by the relative dielectric constant $(\varepsilon_1+\varepsilon_2)/\varepsilon$ of the film. Specifically, if $(\varepsilon_1+\varepsilon_2)/\varepsilon qd\ll\!1$ (relatively thick film), then
\begin{equation}
\omega_p(q)=\omega_p^{3D}\sqrt{2qRI_0(qR)K_0(qR)}\,,
\label{omegap}
\end{equation}
whereas one has
\begin{equation}
\omega_p(q)=\omega_p^{2D}(q)=q\sqrt{\frac{8\pi e^2N_{2D}RI_0(qR)K_0(qR)}{(\varepsilon_1+\varepsilon_2)m^\ast}}
\label{omegap2D}
\end{equation}
if $(\varepsilon_1+\varepsilon_2)/\varepsilon qd\gg1$ (ultrathin film), whose $q$-dependence is different, independent of the material of the film and does show the explicit dependence on the substrate and superstrate dielectric properties.

Figure~\ref{fig2}~(a) shows the ratio $\omega_p/\omega_{3D}$ given by Eq.~(\ref{omegapofq}) as a function of the dimensionless variables $qd$ and $(\varepsilon_1+\varepsilon_2)/\varepsilon$ with the largest possible $R/d\!=\!0.5$. (Reducing the $R/d$ parameter lowers the amplitude but does not change the functional behavior of the graph.) The regimes of the relatively thick and ultrathin films are separated by the vertical plane $qd\!=\!(\varepsilon_1+\varepsilon_2)/\varepsilon$. In both cases, the plasma frequency decreases with $d$ at fixed $q$ and is spatially dispersive ($q$-dependent) at fixed thickness. Figure~\ref{fig2}~(b) shows the contour plot of $\omega_p/\omega_p^{3D}$ one obtains by cutting Fig.~\ref{fig2}~(a) with parallel vertical planes of constant $(\varepsilon_1+\varepsilon_2)/\varepsilon$. With increase of this parameter (the direction shown by the vertical blue arrow) one gradually transitions from the thick film to the ultrathin film regime. At fixed thickness, the plasma frequency spatial dispersion is $\sim\!\sqrt{q}\,$ for thick films and changes to become very close to $\sim\!q\,$ for ultrathin films, thereby making it possible to tune the spatial dispersion on demand not only by varying the material composition of the film but also by adjusting its thickness, $R/d$ ratio, and by choosing the substrates and coating layers appropriately.

\begin{figure}[t]
\epsfxsize=9.3cm\centering{\epsfbox{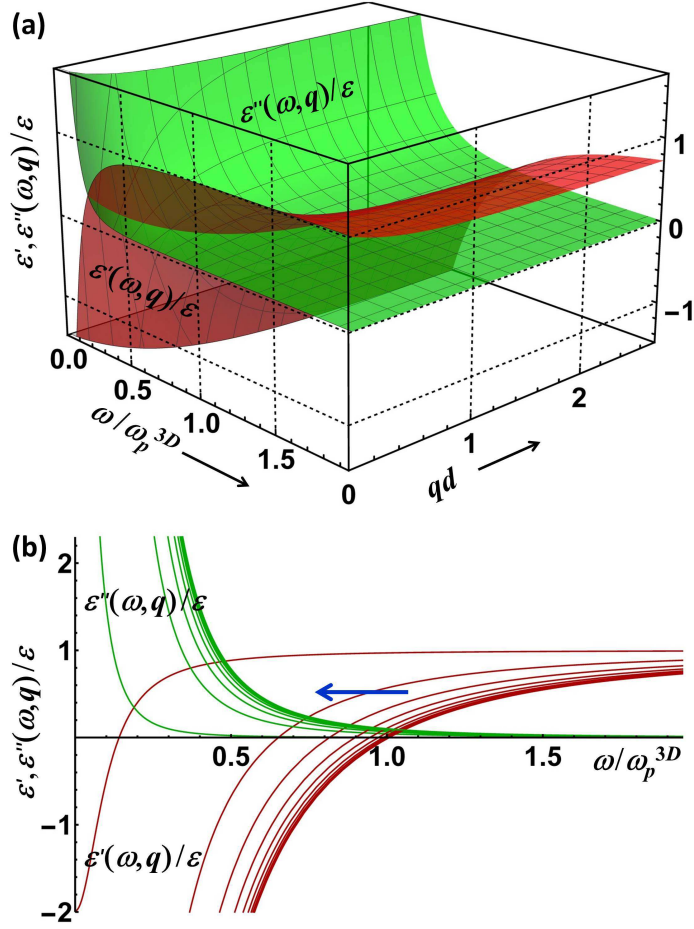}}
\caption{(a)~Real (red) and imaginary (green) parts of Eq.~(\ref{Lindhard}) as functions of the dimensionless variables $\omega/\omega_p^{3D}$ and $qd$. ~(b)~The contour plot one obtains by cutting the graph in (a) with parallel vertical planes of constant $qd$. The thick horizontal blue arrow shows the direction of the $qd$ decrease.}\label{fig3}
\end{figure}

With losses taken into account phenomenologically, the complex-valued frequency-dependent low-momentum dielectric response function in the direction of the cylinder alignment (RPA, commonly referred to as the Drude response function) is of the form
\begin{equation}
\frac{\varepsilon(q,\omega)}{\varepsilon}=1-\frac{\omega_p^{2}(q)}{\omega(\omega+i\gamma)}\,.
\label{Lindhard}
\end{equation}
Here, $\gamma$ is the phenomenological inelastic electron scattering rate and $\omega_p(q)$ is given by Eq.~(\ref{omegapofq}). Figures~\ref{fig3}~(a)~and~(b) show the real ($\varepsilon^\prime\!/\varepsilon$) and imaginary ($\varepsilon^{\prime\prime}\!/\varepsilon$) parts of the dielectric response (\ref{Lindhard}) as functions of the dimensionless variables $qd$ and $\omega/\omega_p^{3D}$. The graphs are calculated for conservative parameter values $\gamma/\omega_p^{3D}\!=0.1$, $(\varepsilon_1+\varepsilon_2)/\varepsilon\!=\!0.1$, and $R/d\!=\!0.5$. In Figure~\ref{fig3}~(b) obtained by cutting the graph in Fig.~\ref{fig3}~(a) with parallel vertical planes of constant $qd$, one can clearly see the approach of $\varepsilon^{\prime\prime}\!/\varepsilon$ to the horizontal axis and the shift of the zero point of $\varepsilon^\prime\!/\varepsilon$ from unity towards values lower than unity as $qd$ decreases (the direction shown by the horizontal blue arrow) to approach the ultrathin film limit. These correspond to the dissipative loss being decreased at a \emph{fixed} frequency and the plasma frequency being red shifted to go lower than $\omega_p^{3D}$ with the film thickness reduction. The red shift of the plasma frequency is accompanied by the gradual increase of the dissipative loss at the \emph{plasma} frequency. These features are similar to those previously reported by Bondarev and Shalaev for isotropic plasmonic films~\cite{BondarevShalaev17}. Here they only manifest themselves in the direction of the cylindrical alignment though, with the overall effect controlled by the film thickness and $R/d$ ratio, by the relative dielectric constant $(\varepsilon_1+\varepsilon_2)/\varepsilon$ and by the \emph{effective} plasma frequency $\omega_p^{3D}$ of the film.

\section{Application to periodically aligned carbon nanotube films}

Aligned carbon nanotube films present a typical example of plasmonic films with cylindrical anisotropy. They are in the process of intensive experimental development~\cite{Kono16NatNano,Abram17NL,Abram17PRL}, with a great potential to become the next generation advanced flexible platform for multi\-functional metasurfaces and nonlinear optical devices with adjustable characteristics on demand. Both thick ($d\!\sim\!200-500$~nm) and relatively thin ($d\!\sim\!30-100$~nm) films of aligned CNs are reported to have unidirectional plasma frequencies $\omega_p\!\sim\!\sqrt{q}\,$ controllable by doping, by the film thickness and by the CN length $L$, with thicker films showing blueshifted plasma frequencies as compared to thinner ones of the same length~\cite{Abram17NL,Abram17PRL}.

The above observations are in full agreement with and thus can be explained and understood in terms of Eq.~(\ref{omegap}) (the thick film regime) and the graphs in Fig.~\ref{fig3}. The plasma frequency $\sqrt{q}$-dependence observed indicates that even the films as thin as $30$~nm are still in the thick film regime. The true thin (or \emph{utrathin}) film regime would manifest itself by the (quasi-)\emph{linear} $q$-dependence of the plasma frequency as given by Eq.~(\ref{omegap2D}). Note that (in addition to doping commonly used) the plasma frequency of the aligned CN films can also be adjusted by varying the CN length due to the trivial fact of $L/2\pi$ being the density of states of the 1D momentum space. According to Eqs.~(\ref{omegap}) and (\ref{omegap2D}), this gives $\omega_p\!\sim\!1/\sqrt{L}$ and $\sim\!1/L$ for the plasma frequency $L$-dependence in the thick and thin film regime, respectively. The former was previously observed experimentally~\cite{Abram17NL}.

Note also that the ultrathin plasma frequency expression (\ref{omegap2D}) is in full consistency with the eigen mode spectrum of individual CNs in free space. Indeed, in general the Green's tensor poles determine the dispersion equation for the eigen modes of the problem, which for the single-wall CN problem has the form~\cite{Bondarev09PRB}
\begin{equation}
\frac{\omega}{\sigma(\omega)}=-4\pi iR\left[q^{2}-(\omega/c)^{2}\right]I_p\!\left[\sqrt{q^{2}-(\omega/c)^{2}}R\right]K_p\!\left[\sqrt{q^{2}-(\omega/c)^{2}}R\right].
\label{disprel}
\end{equation}
Here $p\!=\!0,1,2,...$ and $\sigma(\omega)\!=\!\sigma_{intra}(\omega)\!+\!\sigma_{inter}(\omega)$ stands for the CN dynamical axial surface conductivity (along the nanotube symmetry axis) with the two terms representing intra- and interband transitions, respectively. In the nonrelativistic limit ($c\!\rightarrow\!\infty$) with $p\!=\!0$ (the dominant contribition) Eq.~(\ref{disprel}) takes a simplified form as follows~\cite{Ando2009,GarciaAbajo15}
\begin{equation}
\frac{\omega}{\sigma(\omega)}=-4\pi i\,q^2R\,I_0(qR)K_0(qR)\,,
\label{dispnonrel}
\end{equation}
which brings one to Eq.~(\ref{omegap2D}) with $\varepsilon_1\!=\!\varepsilon_2\!=\!1$ when one ignores the interband electronic transitions whereby $\sigma(\omega)\!\approx\!\sigma_{intra}(\omega)\!\approx\!iN_{2D}e^2/(m^\ast\omega)$ with the relaxation neglected.

Obviously, using the intraband term to approximate $\sigma(\omega)$ is only good for finding plasma frequency modes \emph{far} from the interband transition frequencies of the CN. Interband transitions come from the circumferential quantization associated with the transverse electron confinement on the CN surface~\cite{Ando2005}. They manifest themselves either as excitons in optical spectroscopy or as interband plasmons (of similar but \emph{not equal} excitation energy) in electron energy-loss spectroscopy experiments~\cite{Bondarev09PRB}. In the neighborhood of the CN interband transitions the dispersion equation (\ref{dispnonrel}) generates the plasma frequency modes that are completely different from the (quasi-)~linearly dispersive modes of Eq.~(\ref{omegap2D}). For example, close to an interband transition of energy $E_u$ one has $\sigma(\omega)\!\approx\!\sigma_{intra}(\omega)\!\sim\!2i\hbar\omega/[E_u(\hbar\omega^2\!-E_u^2)]$ with the relaxation neglected~\cite{Ando2005}. Using this in Eq.~(\ref{dispnonrel}) and ignoring the terms of the infinitesimal order higher than linear in $q$, one obtains the nondispersive mode $\hbar\omega_p\approx\!E_u$. Careful analysis with the relaxation included shows that such modes can be found both in metallic and in semiconducting single-wall CNs of not too big diameters ($\sim\!1$~nm) as the resonances of the real part of $1/\sigma(\omega)$, the electron energy-loss response function~\cite{Bondarev09PRB}. Interband plasmons were previously shown theoretically to control many processes of both fundamental and applied importance both in individual pristine and in hybrid CN systems~\cite{Bondarev10JCTN,Bondarev14PRB,Bondarev12PRB,Bondarev16PRB,Bondarev15OE}. Interband plasmons carry the features of individual single-wall CNs and so they should manifest themselves in the \emph{ultrathin} films of aligned small-diameter CNs, both metallic and semiconducting ones, --- the regime that does not seem to have been reached experimentally as yet.

\section{Conclusion}

In this article, the plasma frequency and the dielectric response function are derived and analyzed for the finite-thickness plasmonic film formed by periodic parallel arrays of metallic cylinders embedded in a host dielectric matrix. The plasma frequency of the system is shown to exhibit the unidirectional spatial dispersion $\omega_p\sim\!\sqrt{q}\,$ and $\sim\!q\,$ for the thick and ultrathin films, respectively. The associated \emph{unidirectional} dielectric response nonlocality can be tuned by the film material composition, the film thickness, the cylinder length, the cylinder-radius-to-film-thickness ratio, and by an appropriate choice of substrates and superstrates of the film. The theory developed is discussed in application to the finite-thickness periodically aligned carbon nanotube films for which the importance of the nondispersive interband plasmon modes of individual nanotubes is stressed in the ultrathin film regime. Periodically aligned carbon nanotube film systems are currently in the process of intensive experimental development in the hope of creating a new generation of ultrathin multifunctional metasurfaces and nonlinear optical devices with characteristics adjustable on demand by means of the CN diameter, chirality and periodicity variation~\cite{Kono16NatNano,Abram17NL,Abram17PRL}.

\section*{Acknowledgments}
I.V.B. is supported by the US National Science Foundation grant DMR-1830874. Stimulating and fruitful discussions with Abram L. Falk of the IBM Watson Research Center are gratefully acknowledged.



\end{document}